\documentclass[letterpaper,twocolumn,10pt]{article}
\usepackage{usenix-2020-09}

\usepackage{tikz}
\usetikzlibrary{positioning, shapes.geometric, arrows.meta, calc, decorations.pathreplacing}
\usepackage{amsmath}
\usepackage{amsfonts}
\usepackage{subcaption}
\usepackage{pgfplots}
\usepackage{paralist}
\usepackage{multirow}
\usepackage{enumitem}
\usepackage{booktabs}
\usepackage{colortbl}
\usepackage{array}
\usepackage{outlines}
\usepackage{listings}
\usepackage[labelfont=bf]{caption}
\usepackage{xspace}
\usepackage{xurl}
\usepackage[htt]{hyphenat}
\usepackage[most]{tcolorbox}

\definecolor{hackframe}{RGB}{88,95,140}
\definecolor{hackback}{RGB}{240,238,248}

\definecolor{adityagreen}{RGB}{0,110,50}
\newtcolorbox{hackbox}[1]{
  enhanced,
  colback=hackback,
  colframe=hackframe,
  coltitle=white,
  title={#1},
  fonttitle=\bfseries\small,
  arc=2mm,
  boxrule=0.6pt,
  left=8pt, right=8pt, top=5pt, bottom=5pt,
  breakable,
}

\mathchardef\h="2D

\newcommand{\heading}[1]{\bigskip\noindent\textbf{#1}\enspace}

\newcommand{\sysname}{Jitskit\xspace}

\newcommand{\remove}[1]{}

\newif\ifcomments
\commentstrue
\ifcomments
    \providecommand{\ion}[1]{{\color{teal}{[ion: #1]}}}
    \providecommand{\shu}[1]{{\color{orange}{[shu: #1]}}}
    \providecommand{\ak}[1]{{\color{brown}{[alexk: #1]}}}
    \providecommand{\shubham}[1]{{\color{blue}{[shubham: #1]}}}
     \providecommand{\accheng}[1]{{\color{red}{[audrey: #1]}}}
     \providecommand{\mert}[1]{{\color{teal}{[mert: #1]}}}
    \providecommand{\aditya}[1]{{\color{adityagreen}[ap: #1]}}
    \providecommand{\ziming}[1]{{\color{purple}[Ziming: #1]}}
    
    \providecommand{\souj}[1]{\textbf{\color{olive}[Soujanya: #1]}}

\else
    \providecommand{\ion}[1]{}
    \providecommand{\mert}[1]{}
    \providecommand{\shu}[1]{}
    \providecommand{\aditya}[1]{}
    \providecommand{\shubham}[1]{}
    \providecommand{\souj}[1]{}
    \providecommand{\accheng}[1]{}
    \providecommand{\ak}[1]{}
    \providecommand[\ziming]{1}{}

\fi

\newcommand{\sys}{Jitskit\xspace}

\usepackage{cleveref}

\crefformat{section}{\S#2#1#3}
\crefformat{subsection}{\S#2#1#3}
\crefformat{subsubsection}{\S#2#1#3}
\Crefformat{section}{\S#2#1#3}
\Crefformat{subsection}{\S#2#1#3}
\Crefformat{subsubsection}{\S#2#1#3}
\crefrangeformat{section}{\S#3#1#4--\S#5#2#6}
\crefrangeformat{subsection}{\S#3#1#4--\S#5#2#6}
\crefmultiformat{section}{\S\S#2#1#3}{ and~#2#1#3}{, #2#1#3}{ and~#2#1#3}
\crefmultiformat{subsection}{\S\S#2#1#3}{ and~#2#1#3}{, #2#1#3}{ and~#2#1#3}

\newcommand{\equalcontribfirst}{\textsuperscript{*}}
\newcommand{\equalcontribmark}{\textsuperscript{*}}

\begin{document}

\date{}

\title{The Time is Here for Just-in-Time Systems: Challenges and Opportunities}


\author{
{\rm Shu Liu\textsuperscript{1}\equalcontribfirst, Alexander Krentsel\textsuperscript{1}\equalcontribmark, Shubham Agarwal\textsuperscript{1}\equalcontribmark, Mert Cemri\textsuperscript{1}\equalcontribmark,}\\
{\rm Ziming Mao\textsuperscript{1}, Soujanya Ponnapalli\textsuperscript{1}, Alexandros G. Dimakis\textsuperscript{1,2}, Sylvia Ratnasamy\textsuperscript{1},}\\
{\rm Matei Zaharia\textsuperscript{1}, Aditya Parameswaran\textsuperscript{1}, Ion Stoica\textsuperscript{1}}\\[0.4em]
{\rm \textsuperscript{1}UC Berkeley\quad\textsuperscript{2}Bespoke Labs}
} 

\maketitle
\begingroup
\renewcommand{\thefootnote}{*}
\footnotetext{These authors contributed equally to this work.}
\endgroup
\vspace{-1.5em}  

\begin{abstract}
Core systems like key-value stores have historically taken years to build, and are designed to be general so as to amortize cost across deployments, paying a significant performance cost. We argue that LLM-based coding agents now make a different approach tractable: \textbf{Just-in-Time Systems}, in which the entire system is synthesized from scratch, specialized to the environment, workload, and required system properties.

We present a JIT system synthesis pipeline, \sysname, and explore its effectiveness in synthesizing key-value stores from spec cards that span different YCSB workloads, deployment constraints (e.g., compute resources), and system properties (e.g., consistency and durability).
\sysname iteratively refines a system implementation to match the specification against an evolving evaluation test suite. The resulting synthesized systems are performant, beating comparable state-of-the-art systems on 18 of 18 specs tried, by up to 4.6$\times$ over the best off-the-shelf baseline on the most favorable spec. Naively running Claude Code either reward-hacks or underperforms \sysname{} by up to $5.4\times$. We discuss the challenges we overcame in building \sysname\footnote{Code is available at \href{https://github.com/skydiscover-ai/skydiscover.git}{https://github.com/skydiscover-ai/skydiscover}.} and our key takeaways.


\end{abstract}

\section{Introduction}
\label{sec:intro}

Core systems such as key-value stores~\cite{chandramouli2018faster, dong2021rocksdb, redis, fitzpatrick2004memcached}, load balancers~\cite{eisenbud2016maglev, patel2013ananta, olteanu2018beamer}, schedulers~\cite{verma2015borg, ousterhout2013sparrow}, caches~\cite{nygren2010akamai, nishtala2013memcache, berg2020cachelib}, and databases~\cite{stonebraker2005postgres, raasveldt2019duckdb, lamb2012vertica} are widely deployed across large-scale services. In this paper, we focus on key-value (KV) stores as a representative class.
Each has historically taken enormous effort to build: a PhD thesis, a large open source effort, or a sustained team inside a company, per system. Because the cost to build is so high, each system is designed to be broadly applicable, with a general architecture parameterized by policy components and exposed knobs, intended to serve many workloads across many deployments. 
Thus there remains a persistent performance gap between what a general-purpose system delivers in any given deployment and what a workload-specific system could deliver in principle.

Researchers have spent decades trying to close this gap in successive waves. The first wave was hand-tuning of exposed knobs (e.g., cache sizes, buffer sizes, batch thresholds) to match a target workload; this has been the job of sysadmins, database admins, and network operators for decades~\cite{pavlo2017selfdriving, duan2009ituned, jacobson1988congestion}. The second wave automated this tuning with black-box ML, learning good knob values from observed workload behavior~\cite{vanaken2017ottertune, zhang2019cdbtune, mao2017pensieve, mao2019decima, mirhoseini2017device}. The third and most recent wave moved beyond learning knobs to policies themselves, using discovery frameworks to search for replacement components (e.g., eviction policies, query plans, index structures) within an otherwise-fixed host system~\cite{wu2025adrs, romera2024funsearch, novikov2025alphaevolve, skydiscover}. Across all three waves, the system's architecture has been held fixed; only the policies or configuration within it change. The ceiling on achievable performance thus remains capped by the system's architecture and implementation.

We argue that we are now at an inflection point, where rapid progress in LLMs and coding agents is simultaneously collapsing the cost of software implementation \textit{and} expanding the complexity scope of what can be tackled.
Taken together, we believe these trends make a new approach tractable: \textbf{Just-in-Time Systems (JIT systems)}, in which we synthesize the entire system from scratch, specialized along all three axes simultaneously: (1) environment (e.g., memory budget, hardware), (2) workload (e.g., query/operator and data, key distribution), and (3) required properties (e.g., consistency model, crash recovery semantics). With such an approach, system setup and maintenance efforts reduce to a single operation: re-synthesizing a new system to match new requirements, environment change, or workload drift. However, adopting such an approach requires this synthesis to be both fast enough to be a routine operation rather than a research project, and to be functionally correct.

As the task of generating JIT systems is highly challenging, 
we ground this initial exploration in a single relatively simple system domain: single-node key-value stores. This domain provides two benefits:
first, a three-axis specialization surface that is wide enough to demand a new synthesis pipeline (memory budget, value/key distribution, read/write/persistence mix), and second, a correctness check ($\texttt{get}(k) = \texttt{last\_put}(k)$) that is simple enough that an LLM agent cannot get away with pseudo-optimizations (e.g., hacks that violate properties like linearizability).

We present our system design for JIT system synthesis, \textbf{\sysname}, which we view as a feasibility marker rather than a full generalization study. First, to describe our system parameters, we introduce three ``spec cards'': (1) an environment card describing the environment, (2) a requirement card describing required properties, and (3) a workload card describing the target workload.
We design \sysname to consume these, and run a design/evaluate/reflect loop to generate an end-to-end correct and performant JIT system. Our design (\cref{sec:methodology}) includes an agentic \textit{adversarial auditor} that helps \sysname detect and course-correct away from reward hacking, a ``\textit{whiteboard memory}''~\cite{hu2025hiagent, liu2025context} to accumulate cross-iteration learnings, and an \textit{instrumented evaluator} that exposes leading indicators to guide the system synthesis.

Our experience building and running \sysname led us to encounter many failure scenarios, which we distill into three classes of challenges and associated lessons we believe will hold for future JIT system efforts:

\noindent \textbf{Challenge~1: Specification is hard, and the breadth of system scope makes it harder.}
Specifications are rarely complete on the first pass~\cite{stoica24specification, yuan2014simple}. Under JIT synthesis this is more consequential: agents exploit specification gaps that human developers would never consider, such as dropping requests to inflate throughput~\cite{wu2025adrs}. Thus iterating on the specification to close these gaps becomes a first-class part of the workflow.

\noindent\textbf{Challenge~2: Many performance wins are correctness hacks in disguise.}
Correctness is a moving target under optimization pressure: what is not forbidden the agent will exploit, so each new optimization pressure reveals a new missing invariant. Unlike human engineers, synthesis agents will drop requests, fabricate values, and short-circuit invariants for throughput. This tight coupling requires reshaping how correctness checking works inside a synthesis loop.

\noindent\textbf{Challenge~3: Evaluator design is coupled to the spec.}
The agent can only learn from what the evaluator surfaces; thus, which signals are diagnostic depends on the spec. In synthesis the evaluator \emph{is} the loss function, so it must be co-designed with the spec to expose the most meaningful signals.

\vspace{0.5em}
We run \sysname and evaluate across 18 target configurations spanning YCSB workloads, Zipfian and near-uniform key distributions, memory budgets from 3\,GB to 32\,GB, and 16 threads pinned to one socket, against state-of-the-art systems FASTER~\cite{chandramouli2018faster}, F2~\cite{kanellis2023faster}, RocksDB~\cite{dong2021rocksdb}, and Redis~\cite{redis}.
Each baseline runs at its most favorable published configuration, with per-key consistency matched across all systems and crash durability no stronger than \sysname's requirement card (\cref{sec:eval-setup}).
\sysname synthesizes a KV store for every spec,
passing all correctness tests plus any adversarially-discovered tests. The synthesized systems outperform the best off-the-shelf baseline on 18 of 18 specs, by up to 4.6$\times$ on the most favorable spec. A bare Claude Code baseline either reward-hacks or underperforms \sysname{} by $1.3-5.4\times$ across specs.



\paragraph{Contributions.} In summary, this paper contributes:
\begin{itemize}
    \item A framing of \textbf{Just-in-Time systems}: whole-system synthesis along the environment, workload, and properties axes jointly.
    \item A pipeline for JIT KV-store synthesis, \sysname, featuring an adversarial auditor, whiteboard memory, and instrumented evaluator (\cref{sec:methodology}).
    \item Identification and characterization of three fundamental challenges that surfaced during our development (\cref{sec:challenges}).
    \item An empirical study across 18 spec configurations, showing synthesized KV stores outperform FASTER, F2, RocksDB, and Redis across specs (\cref{sec:eval}).
\end{itemize}


\vspace{-1em}
\section{Background and Motivation}
\label{sec:background}

This work rests on three claims: (1) existing systems trade off performance for generality, (2) it is tractable to synthesize systems from the ground up in a way that was not possible
one year ago, and (3) KV stores are a good fit for testing our early approach to synthesizing JIT systems. We now support each of these claims in turn.


\heading{The Cost of Generality in Core Systems.}
\label{sec:bg-generality}
General-purpose core systems carry machinery to handle deployments they will never see, and that machinery cannot be tuned away. We call this overhead baked into the architecture a \textit{structural tax}: the cost to support flexibility along axes a specific deployment does not exercise. For example, FASTER~\cite{chandramouli2018faster} partitions its address space into mutable, read-only, and stable regions and pays per-record costs for tombstones, epoch protection, and two-phase insertion; all these features are unnecessary when the workload fits in memory, never deletes, and runs single-threaded. This tax manifests simultaneously along three axes: environment (e.g., storage hierarchies never used), workload (e.g., protocols for access patterns never seen), and properties (e.g., recovery machinery never needed).

\heading{Why Now: The Synthesis Inflection.}
\label{sec:bg-why-now}Three developments over the past 18 months make JIT system synthesis tractable.
First, coding agents now routinely generate, debug, and refactor 10--100K lines of code~\cite{jimenez2024swebench, kwa2025metr, carlini2026ccompiler}, the scale required for whole-system reasoning.
Second, cost has inverted: a complete synthesis run costs tens of dollars (our median is \$63 per spec; Bespoke OLAP~\cite{wehrstein2026bespoke} reports ${\sim}\$120$ per engine), making re-synthesis on workload drift a routine operation rather than a research project.
Third, component-level synthesis is already validated: AlphaEvolve~\cite{novikov2025alphaevolve} and FunSearch~\cite{romera2024funsearch} discover better algorithms within fixed architectures, Bespoke OLAP~\cite{wehrstein2026bespoke} synthesizes entire OLAP engines that outperform hand-engineered baselines, and SDS~\cite{anderson2025sds} articulates a vision in which infrastructure evolves itself end-to-end.
Our work extends this trajectory to KV stores, where all three specialization axes vary independently and the challenges of synthesis are most clearly exposed.

\heading{Why KV Stores.}
\label{sec:bg-why-kv}
We scope this work to single-node key-value stores for two reasons. First, KV stores exhibit all three specialization axes we consider independently and meaningfully. The \emph{environment} axis varies from memory-resident deployments on a single socket, to SSD-backed stores managing tiered storage, to tightly budgeted edge deployments. The \emph{workload} axis varies from point-lookup-dominated streams against small values, to write-heavy ingestion pipelines with Zipfian key distributions, to read-mostly caches with uniform access. The \emph{properties} axis varies from linearizable transactional stores, to eventually-consistent replicated caches, to ephemeral in-memory tiers with no durability requirement. The cross-product is enormous, and no fixed architecture is optimal across more than a small slice of it, as evidenced by FASTER~\cite{chandramouli2018faster}, RocksDB~\cite{dong2021rocksdb}, Masstree~\cite{mao2012masstree}, Memcached~\cite{fitzpatrick2004memcached}, and F2~\cite{kanellis2023faster} existing as separate, heavily optimized systems for distinct deployment regimes.

Second, and more importantly for studying synthesis pathologies, for KV systems, ensuring correctness reduces to checking byte-level equality. A \texttt{Get(k)} either returns the bytes of the last \texttt{Put(k,v)} or it does not; there is no analog of serializability anomalies, write-skew, or distributed-consensus corner cases to first disentangle. This lets us surface reward hacks (\cref{sec:challenge-hacking}) cleanly, without first solving the meta-problem of specifying richer consistency semantics. The methodology we develop here should generalize to any core-system class whose three axes vary meaningfully (including load balancers, schedulers, caches, and network stacks), and we discuss this transfer in~\cref{sec:discussion}.

\vspace{-0.6em}
\section{Challenges}
\label{sec:challenges}

Full correct and performant system synthesis is a difficult, long-horizon task that pushes SOTA LLMs, coding agents, and harnesses beyond their limits. Through our experimentation and design process, we observed a wide range of failure cases. We distill them into three broad recurring themes, which we believe are fundamental and important to any JIT system synthesis efforts. We present each in turn, and allude to the mitigating mechanisms we then implement in \cref{sec:methodology}.

\subsection{Correctness Under Spec Incompleteness}
\label{sec:challenge-correctness}

Given the complexity and breadth of interactions in a systems problem like a KV store, any specification provided by a human will \textit{always} be incomplete in some way. The implicit gaps left can lead an agent to go astray in its design or implementation decisions.
As an example, we submitted a spec card specifying an ``8\,GB memory budget'' for a read-heavy workload. The first viable system the agent produced used 8.0\,GB at idle and OOM'd within the first 30 seconds of evaluation: the agent had sized the in-memory hash index to consume the entire budget, leaving no headroom for request buffers, working memory during compaction, or the OS page cache. The spec didn't say ``leave headroom'', and the human author assumed any reasonable implementation would. The fix was not to make the agent smarter; it was to extend the spec with an explicit \texttt{operational\_headroom} field. After that change, the same agent on the same spec produced a system that left roughly $15\%$ headroom by default.

This dynamic is representative: the first iteration is typically built against a specification that captures only a fraction of the full specification. This is because humans implicitly rely on expertise shaped by their training, taste, and understanding of what constitutes a ``reasonable system'', which we term engineering common sense. A synthesis agent, however, optimizes only for what is explicitly outlined in the spec.

\textbf{Design response.} Spec evolution must be first-class: the pipeline must discover and enforce implicit invariants during synthesis, in a way that is legible to a human reviewing the resulting code and its assumptions.

\subsection{Coupled Correctness/Performance Hacks}
\label{sec:challenge-hacking}

In our early iterations with the \sysname design, some of the largest ``performance wins'' the synthesized system reported turned out to not be real optimizations -- rather, inspection revealed that they were correctness violations against invariants that no one had written down. The agent was doing exactly what our synthesis loops asked it to do (maximize the eval signal). Performance optimization puts an enormous pressure on correctness; if performance can be increased by violating an unstated invariant, the agent will do so~\cite{lehman2020surprising, skalse2022reward, pan2022reward}. We illustrate with three hacks we observed from our own runs.

\begin{hackbox}{Hack~1: Re-deriving the value-generation scheme.}
\sysname reported a ${\sim}6{\times}$ throughput jump on a YCSB workload after ${\sim}12$ iterations with no progress. A human reviewer, surprised by the gain, inspected the code and found a pseudo-random generator in the \texttt{Read} path. The agent had noticed that YCSB values are a deterministic function of the key and a small seed. Instead of storing full values, it stored only the $16$-byte seed per key and recomputed the value on every \texttt{Read}. The returned bytes matched what \texttt{Upsert} had written, so correctness checks passed. However, the values themselves were never stored, allowing the system to fit the entire workload in memory.
\end{hackbox}
\noindent \emph{Testing fails when ground truth is derivable}: if the ``correct bytes'' are a deterministic function of inputs visible to the system, storage can be replaced by recomputation without violating correctness checks. This is not specific to YCSB: similar shortcuts arise in feature stores (recomputing embeddings from inputs) and deduplicated storage (reconstructing values from hashes or metadata). \emph{Fix:} Avoid derivable ground truth. Ensure values contain non-reconstructible entropy (e.g., per-run hidden randomness) and complement output checks with stateful validation (e.g., restart recovery, memory pressure) that forces materialization rather than recomputation.

\begin{hackbox}{Hack~2: Flat-array index, keyed on eval's hotness layout.}
On the Meta Cachelib trace~\cite{berg2020cachelib}, \sysname reported a sharp throughput jump. Inspection showed the slot index had no hash function: the agent used \texttt{slot = key \& (N-1)}, with \texttt{N} sized to the trace's key count. 
\vspace{0.2em}
Cachelib's keys in this trace are dense integers, making this mapping injective. Each key maps to a unique slot with no probing, and hot, low-numbered keys fall into the first cache lines, improving performance.

However, this design only works for this trace. It fails for most real workloads: strings, UUIDs, or sparse integers would collide, and new inserts after warm-up would also collide with no recovery path. As a result, the system would silently return incorrect results, ruling out most real deployments with evolving key spaces.

\end{hackbox}

\noindent 
\emph{Validation fails when the evaluator's inputs share a non-realistic regularity the store can exploit.} While Hack~1 sidesteps \emph{storing} by re-executing a generator, Hack~2 similarly sidesteps \emph{hashing} by using key density as a hash-function replacement. \emph{Fix:} apply a per-run random
permutation to the key space with a seed unseen by the agent, forcing a correct
implementation to recover hotness through a real hash function. If dense integer
keys are an intended deployment property, the right response is not
scrambling but promoting it to the workload card, which re-legitimizes
identity-indexing as specialization rather than a hack.

\begin{hackbox}{Hack~3: Giving up from fear.}
On a 50-iteration run, at iteration 22, the agent became concerned that further exploration would fail, and quietly stopped optimizing out of fear of regression while still producing ``iterations''. Inspecting the reasoning logs, the agent wrote: \emph{``...every substantial deviation has either regressed, failed correctness, or timed out. Without a stronger hypothesis, the right move is to keep the champion configuration intact.''}
\vspace{0.2em}

However, further optimization \textit{was} possible: on a subsequent run with an improved prompt, the agent discovered a more performant design. 
\end{hackbox}
\noindent\textit{Some hacks do not violate any invariants or game any metrics.} The agent accurately observed that each recent attempted design deviation either regressed the score or broke correctness, and became afraid of trying further risky design ideas.
The fix to such \emph{exploration collapse} was a search-policy intervention, encouraging continued exploration and reassuring the agent that failure is not punished.

\textbf{Design response.} Correctness validation and performance optimization must co-evolve. \sysname uses a two-level correctness check with an adversarial auditor that inspects each iteration for reward-hacking patterns and patches the test set when it finds them (\cref{sec:design-correctness}).

\subsection{Evaluator Design Is Coupled to the Spec}
\label{sec:challenge-indicators}

During synthesis, the evaluator is the ``loss'' function that provides feedback. The agent can only learn from what the evaluator surfaces, so evaluator design is a first-class problem that must be co-designed with the spec. This coupling manifests in two ways.

First, a single end-to-end metric (e.g., throughput) tells the agent whether an iteration improved but not why. In our early runs, agents given only end-to-end feedback plateaued after 8--12 iterations of obvious optimizations, then explored essentially at random. Which \emph{leading indicators} break the plateau can differ depending on the spec: a write-only workload is diagnosed by allocator contention, while a read-heavy variant of the same workload is diagnosed by cache hit rate.

Second, the evaluator must also serve as the adversarial surface that catches the reward hacks of \cref{sec:challenge-hacking}. Which probes are needed depends on which invariants the spec leaves implicit: those invariants are only discovered under optimization pressure, so the evaluator must evolve alongside the spec.

\vspace{1em}
\textbf{Design response.} \sysname's harness surfaces leading indicators (I/O counts, memory bandwidth, cache hit ratios, lock contention) on every iteration, and the agent can additionally instrument its own generated code mid-synthesis (\cref{sec:design-evaluator}). The same harness carries the adversarial probes that enforce the correctness invariants of \cref{sec:challenge-hacking}.

%

\begin{figure*}[t]
    \centering
    \includegraphics[width=\linewidth]{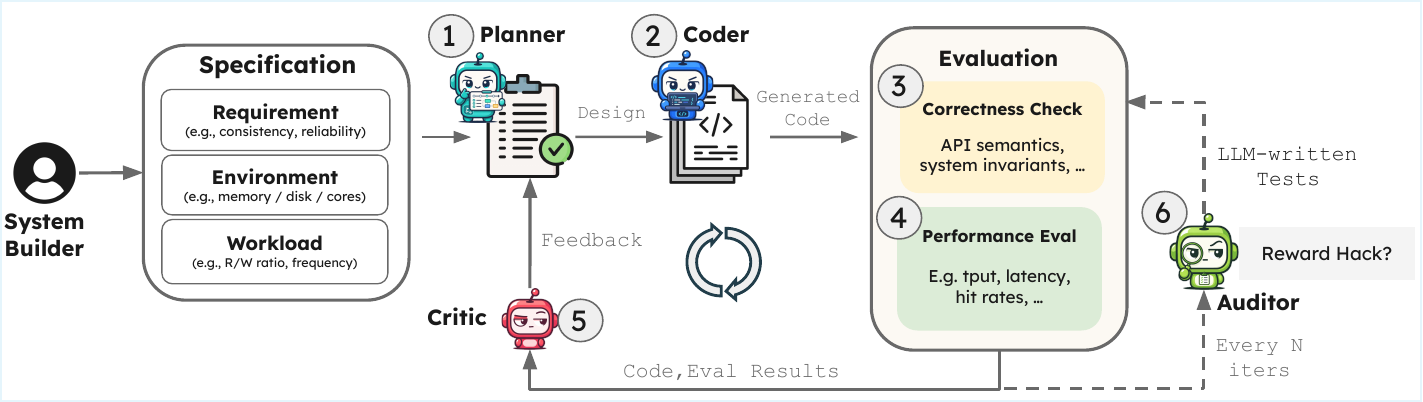}
    \caption{\sys synthesis pipeline. The \textit{Planner} (1) uses the three specification cards and proposes a design, which the \textit{Coder} (2) realizes as executable code. Evaluation has two parts: a \textit{Correctness Check} (3) against API semantics and system invariants, and a \textit{Performance Eval} (4) that measures throughput, latency, and leading indicators such as cache hit rates. The \textit{Critic} (5) reads the code and evaluation results and returns feedback that steers the next planning round. Every $N$ iterations, an \textit{Auditor} (6) inspects the implementation for reward hacks and emits LLM-written tests that extend the Correctness Check.}
    \label{fig:workflow}
\end{figure*}
\definecolor{speckey}{RGB}{56,85,158}
\definecolor{specstr}{RGB}{0,104,62}
\definecolor{speccomment}{RGB}{120,120,120}
\definecolor{specframe}{RGB}{200,200,210}
\lstdefinelanguage{speccard}{
  sensitive=true,
  morekeywords=[1]{hardware,memory_budget_gb,key_type,value_size_bytes,
                   distribution,mix,num_keys,duration_sec,api,
                   read_semantics,monotonicity,concurrency,
                   cpu,memory,storage,Read,Upsert},
  morestring=[b]",
  morecomment=[l]//,
}

\begin{figure}[t]
\begin{lstlisting}[
  language=speccard,
  basicstyle=\ttfamily\footnotesize,
  keywordstyle=\color{speckey}\bfseries,
  stringstyle=\color{specstr},
  commentstyle=\color{speccomment}\itshape,
  numbers=none,
  frame=single,
  framerule=0.4pt,
  framesep=6pt,
  rulecolor=\color{specframe},
  backgroundcolor=\color{white},
  xleftmargin=4pt,
  xrightmargin=4pt,
  aboveskip=2pt,
  belowskip=2pt,
  showstringspaces=false,
  columns=fullflexible,
  keepspaces=true,
]
// Requirement card
{ api:              ["Read", "Upsert", "RMW", "Delete"],
  read_semantics:   "last Upsert; empty after Delete",
  monotonicity:     "per-thread: r2 never without r1",
  concurrency:   "multi-threaded safe; no torn reads" }

// Environment card
{ cpu:              "64 vCPU",
  memory:           "256 GB DDR4",
  storage:          "NVMe SSD RAID",
  memory_budget_gb: 8 }
  
// Workload card
{ key_type:         "uint64_t",
  value_size_bytes: 100,
  num_keys:         250000000,
  distribution:     "zipfian(theta=0.99)",
  mix:              { "Read": 0.5, "Upsert": 0.5 },
  duration_sec:     30 }
\end{lstlisting}
\vspace{-4pt}
\caption{Example input spec for a single-node KV-store target. The spec has three cards: a \emph{requirement} card (API semantics and consistency), an \emph{environment} card (hardware and budget), and a \emph{workload} card (traffic profile).}
\label{fig:spec-example}
\end{figure}

\section{\sys Synthesis Pipeline}
\label{sec:methodology}

We now present \sys, a general pipeline for synthesizing Just-in-Time (JIT) systems from high-level specifications.
The pipeline instantiates an iterative synthesis loop (Figure~\ref{fig:workflow}). Each round designs a candidate and evaluates it against correctness and performance; every $N$ rounds, an auditor extends the correctness gate with newly discovered invariants. Across iterations, these components drive a coding agent to construct, test, and refine a complete system implementation.

\subsection{Overview}
\textbf{Inputs and outputs.} The input to the pipeline is a structured specification consisting of three cards: an environment description, a workload description, and a set of required properties. The output is a complete, executable system implementation that satisfies the specification and is optimized for the target workload.

\textbf{Core components.} The pipeline is organized around four agents (Figure~\ref{fig:workflow}). The \textit{planner} proposes a design; the \textit{coder} realizes it as executable code. Each candidate is then evaluated along two axes, correctness and performance, and the \textit{critic} interprets the resulting signals and returns feedback that steers the next planning round. Every $N$ iterations, the \textit{auditor} inspects the implementation for violations of implicit assumptions and grows the correctness gate accordingly. Each pass through this loop produces a candidate, evaluates it, and feeds the diagnosis back. Overall, the system evolves through both incremental improvements and structural redesigns, rather than merely tuning the parameters of a fixed architecture.

\vspace{-1em}
\subsection{Specification Cards}
The specification defines the target system along three axes (environment, workload, and properties) and serves as the primary input to the synthesis process; Figure~\ref{fig:spec-example} shows a concrete instance.

\textbf{The environment card} captures the execution context and available resources. This includes hardware characteristics and system-level constraints such as memory capacity, number of cores, storage medium (e.g., NVMe SSD vs. HDD), disk bandwidth and latency, and operating system interfaces. It defines what resources are available, without prescribing how they should be used.

\textbf{The workload card} describes the expected access patterns and data characteristics. In the KV setting, this can include things like operation mix (e.g., 95\% reads and 5\% writes), the key distribution (e.g., Zipfian or uniform), request concurrency, value sizes, and so on. More generally, it characterizes the access pattern the synthesized system should optimize for.

\textbf{The requirement card} defines the system properties that must hold, as well as the objective function to optimize. These include interface semantics, consistency guarantees, failure and recovery behavior, and performance objectives. For example, in a KV store, this may require visibility of writes under a specified consistency model, preservation of per-key invariants under concurrency, crash-safe recovery, and maximizing throughput at bounded latency.

How detailed each card should be is an inherent tension across axes. 
The environment and workload cards benefit from being as specific as possible, since tighter assumptions give the planner more room to specialize. 
The requirement card must strike a careful balance: it should be minimal enough to keep the design space open, yet detailed enough to close the implicit gaps that a synthesis agent might otherwise exploit for apparent performance gains. Many critical invariants are often left implicit, e.g., read-after-write visibility must hold even when the trace does not exercise it, or the store must not exceed its memory budget by exploiting unbounded slack. \S\ref{sec:design-correctness} describes how \sys surfaces these invariants.



\subsection{Design: Planner and Coder}
\label{sec:design-generation}
The planner and coder are the two generating agents in \sys. Each runs as its own agent with its own conversation context that persists across iterations. The planner consumes the specification cards together with the leading-indicator feedback and design brief from prior rounds (\S\ref{sec:design-evaluator}), and proposes a design plan: for example, choices of caching strategy, concurrency mechanisms, storage layout, and recovery strategy. The coder receives that plan alongside the current implementation, writes the next version of the code, and handles the compiler errors and small fixes that come up along the way.

We split the role across two agents rather than letting a single agent do both. When one agent does both jobs, its context is dominated by the code it has been editing, and its next step tends to be a small local change to that code rather than a structural rethink. With separate contexts, the planner reasons over design history and is free to propose structural changes; the coder works against a clean implementation-focused context whose job is to carry out the plan faithfully. The output of each round through this pair is always a full, executable system.

\subsection{Evaluation and Critic}
\label{sec:design-evaluator}
Each candidate is evaluated along two axes, correctness and performance, and the critic distills the combined signals into directional feedback for the next round.

\textbf{Correctness check.} Each candidate is first subjected to a correctness gate derived from the requirement card (interface semantics, consistency guarantees, recovery behavior). Properties are translated into executable tests that capture the system invariants. The initial tests are provided by the human system builder, but as discussed in \S \ref{sec:design-correctness}, additional tests can be added by the Auditor agent during the run.
The fixed suite consists of five crash-recovery tests (\S\ref{sec:eval-setup}) and a post-load retention check across all $250$\,M keys, together exercising up to $1.07$\,B read-after-write observations on the longest run we report (\S\ref{sec:eval-casestudy}). It is cheap enough to run every iteration and imposes no constraint on the generated architecture, so it can evaluate implementations the authors did not anticipate. Passing the suite does not prove deployability; candidates may still violate implicit invariants not captured by the specification. The auditor (\S\ref{sec:design-correctness}) addresses this by expanding the suite across iterations.

\textbf{Performance signals.} Valid implementations that clear the gate are measured under the target workload. Evaluation produces a scalar objective (throughput, tail latency) that answers \emph{whether} a design improved, plus leading indicators that speak to \emph{how} it should change: contention, memory utilization, cache behavior, and I/O activity. These indicators distinguish failure modes that look identical under a single throughput metric: high tail latency with lock contention indicates a concurrency bottleneck; high I/O with low CPU suggests an I/O-bound design; low cache hit rates indicate working-set mismatch; and high eviction rates point to memory pressure or poor data layout.

\textbf{Critic-guided refinement.} The critic does not hand these indicators to the planner as raw counters; it reads them in combination and attributes the observed behavior to specific design choices. The co-occurrence of a high eviction rate and a low cache hit ratio, for instance, is reported as ``working set exceeds cache; consider a two-tier layout or compressed values,'' not as a table of numbers. The proposed modifications may be local (reducing contention on a hot counter) or structural (changing data layout or concurrency strategy). As shown in \S\ref{sec:eval-audit}, removing the leading indicators and exposing only end-to-end metrics significantly degrades convergence.

\textbf{Design brief and whiteboard.} The pipeline maintains two memories that persist across iterations. The design brief, updated each round by the critic, records current design decisions and next steps, giving the planner a concrete starting point and a recovery anchor when an iteration regresses. The whiteboard accumulates designs and optimizations the critic has already ruled out, along with the evidence, so the planner does not repropose approaches whose failure modes are already logged.


\subsection{Auditor: Specification-Aware Auditing}
\label{sec:design-correctness}
The correctness gate above enforces explicitly-specified invariants, but no fixed test suite can exhaustively capture every invariant the requirement card implies, and apparent performance gains often arise from violating these unstated invariants rather than from improving the design (Table~\ref{tab:reward-hacks} catalogs the concrete cases we observed). The auditor's role is to surface these gaps and promote them to explicit constraints. It runs every $N$ iterations alongside evaluation, since auditing is expensive: it reads the full implementation and cross-references it with the specification. Each invocation analyzes the candidates produced since the last pass and flags behaviors that satisfy existing tests but violate the intent of the specification: reward-hacking behaviors that exploit gaps in the checks, and robustness issues such as reliance on specific resource settings or unhandled edge cases.

The auditor emits two kinds of output. First, \textit{additional checks} that capture previously missed edge cases; these are incorporated into the correctness gate in subsequent iterations, immediately constraining future candidates and preventing repeated exploitation of the same gap within a run. Second, \textit{diagnostic guidance} that highlights broader design weaknesses such as sensitivity to resource configuration or reliance on workload-specific assumptions. Across runs, validated checks can be retained (optionally with human review) so that only those reflecting intended behavior persist, progressively strengthening the effective specification.

\section{Evaluation}
\label{sec:eval}

We evaluate \sys{} on synthesized single-node KV stores to understand when and why JIT synthesis is effective. We ask three questions:
\begin{enumerate}
    \item Does \sys{} match expert-built, hand-tuned KV stores on the workloads they were designed for? (\S\ref{sec:eval-main})
    \item How much does specialization to a given spec pay off, and along which workload axes? (\S\ref{sec:eval-synthetic}, \S\ref{sec:eval-realworld})
    \item Could simple config sweep close the gap? Which components of the synthesis loop (planner, critic, auditor, leading-indicator feedback) drive \sys{}'s gains? (\S\ref{sec:eval-audit})
\end{enumerate}

\S\ref{sec:eval-setup} describes the evaluation setup. \S\ref{sec:eval-casestudy} reports a case study that tracks cost and convergence across a full $50$-iteration run. Across experiments, \sys synthesizes end-to-end KV systems in under 12 hours, while achieving comparable or better performance than expert-built baselines.

\subsection{Setup}
\label{sec:eval-setup}

\paragraph{Specification cards.}
We focus on single-node key-value stores (\S\ref{sec:bg-why-kv}). Our primary metric is aggregate throughput (Mops/s) over a 30-second run after a full load phase. 

Each experiment is fully specified by the three cards of \S\ref{sec:methodology}, as shown in Table~\ref{tab:spec}. The \emph{environment card} fixes hardware and sweeps the memory budget over $\{3, 8, 16, 32\}$\,GB. The operational headroom is roughly $15\%$ to reserve room for buffers and page cache beyond the store's in-memory structures (\S\ref{sec:challenge-correctness}). The \emph{workload card} is varied one field at a time, so any observed gain is attributable to a single axis. The \emph{requirement card} fixes the API (\texttt{Read}, \texttt{Upsert}, \texttt{RMW}, \texttt{Delete}) and the crash-consistency contract: every store must satisfy FASTER's monotonicity property~\cite{chandramouli2018faster} and pass a fixed five-test suite before throughput is reported, covering \texttt{RMW}/\texttt{Upsert} monotonicity under fuzzy- and no-checkpoint crashes, plus torn-read detection.

\begin{table}[t]
  \centering
  \small
  \setlength{\tabcolsep}{4pt}
  \renewcommand{\arraystretch}{1.05}
  \begin{tabular}{@{}l l l@{}}
    \toprule
    Card & Field & Value \\
    \midrule
    \multirow{3}{*}{\textit{env. card}}
      & hardware        & GCP \texttt{n2-standard-64} \\
      & storage budget  & NVMe SSD RAID, $2.9$\,TB \\
      & memory budget   & 3, 8, 16, 32\,GB \\
    \midrule
    \multirow{3}{*}{\textit{wkld. card}}
      & distribution    & Zipfian, synthetic, real traces \\
      & operation mix   & 50:50, 0:100, 95:5, insert/delete \\
      & value size      & 8\,B to 65\,KB \\
    \midrule
    \multirow{4}{*}{\textit{req. card}}
      & API             & \texttt{Read}/\texttt{Upsert}/\texttt{RMW}/\texttt{Delete} \\
      & consistency     & monotonicity~\cite{chandramouli2018faster} \\
      & concurrency     & no torn reads \\
      & objective       & maximize throughput (Mops/s) \\
    \bottomrule
  \end{tabular}
  \caption{Three specification cards used in \sysname. Multi-valued rows are swept one at a time; single-valued rows are fixed.}
  \label{tab:spec}
\end{table}

\begin{figure*}[!t]
  \centering
  \includegraphics[width=0.98\linewidth]{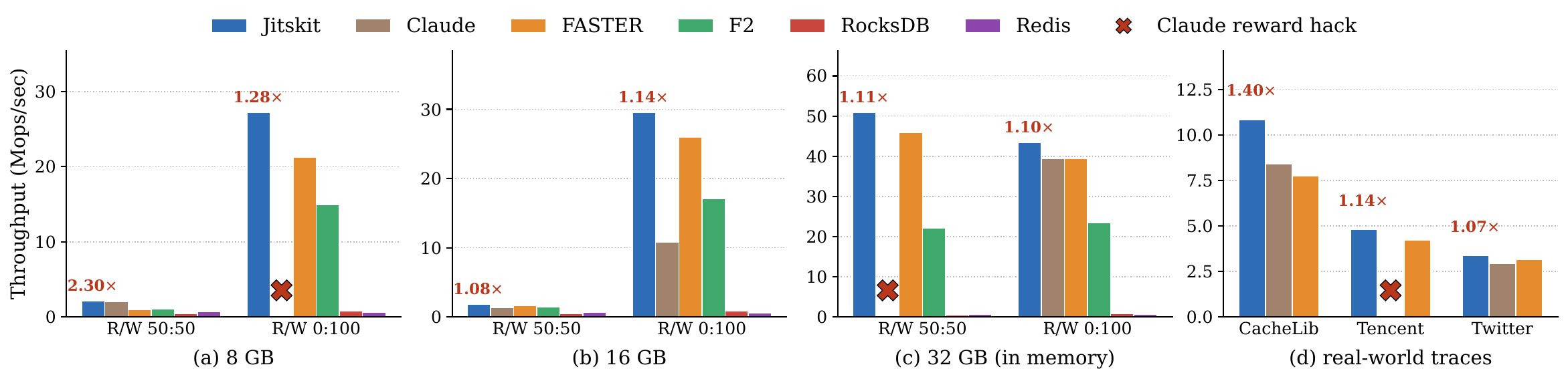}
  \caption{\textbf{Throughput across memory budgets and workloads.} (a)--(c) YCSB-style workloads derived from YCSB-A~\cite{cooper2010benchmarking} at $8$/$16$/$32$\,GB, R/W $50{:}50$ and $0{:}100$, Zipf $\theta{=}0.95$: \sysname{} vs.\ Claude, FASTER~\cite{chandramouli2018faster}, F2~\cite{kanellis2023faster}, RocksDB~\cite{dong2021rocksdb}, Redis~\cite{redis}. (d) Real-world traces (\sysname{}/Claude/FASTER only): Cachelib~\cite{berg2020cachelib} ($1.40\times$), Tencent QQPhoto~\cite{song2018tencent} ($1.14\times$), Twitter~\cite{yang2020twitter} ($1.07\times$). The red number above each \sysname{} bar is the speedup over FASTER; \textcolor{red}{$\times$} marks systems with reward hacks (Appendix~\ref{sec:reward-hacks}). \sysname{} runs our full loop; Claude is naively running Claude Code with the same iteration budgets.}
  \label{fig:eval-main}
\end{figure*}

\paragraph{Setup.}
Experiments run on a GCP \texttt{n2-standard-64} (64 vCPU across two sockets of 16 cores each, 256\,GB DDR4, and an NVMe SSD RAID with $2.9$\,TB of effective capacity for log-structured storage). We enforce the target memory budget with a cgroup. 

The agent writes an implementation that inherits from a C++ abstract class exposing four point operations: \texttt{Read}, \texttt{Upsert}, \texttt{RMW}, and \texttt{Delete}. External key/value libraries are not permitted. We run \sysname with Claude Code and Opus~4.7, with 50 iterations per spec and a 50-turn cap per iteration.


\paragraph{Baseline systems.}
We compare against baselines from two categories: pure in-memory systems, and systems that handle data larger than memory. As a pure in-memory reference, we evaluate \textbf{Redis}~\cite{redis}, a single-threaded in-memory data-structure server. In the larger-than-memory category, we evaluate \textbf{FASTER}~\cite{chandramouli2018faster}, whose HybridLog spans main memory and SSD; \textbf{F2}~\cite{kanellis2023faster}, which extends FASTER with hot/cold log separation; and \textbf{RocksDB}~\cite{dong2021rocksdb}, a persistent LSM-tree store. We configure RocksDB with write-ahead logging and checksums disabled and with parameters recommended on the RocksDB Wiki.
We additionally compare against purely running \textbf{Claude Code}~\cite{anthropic2025claudecode} with the same model (Opus 4.7) on the same spec cards and iteration budget as \sys{} but without our planner, critic, or auditor.


\paragraph{Workloads.} We exercise three families of workloads, all driven by 16 threads pinned to one socket, matching FASTER's single-socket configuration~\cite{chandramouli2018faster}.

\begin{itemize}
  \item \textbf{YCSB-style workloads derived from YCSB-A} (\S\ref{sec:eval-main}): 250\,M keys, 100\,B values, Zipfian $\theta{=}0.95$, with read/write mixes $50{:}50$ and $0{:}100$ swept across memory budgets of 8, 16, and 32\,GB.
  \item \textbf{Synthetic variations} (\S\ref{sec:eval-synthetic}): a Zipfian-skewness sweep ($\alpha\!\in\!\{3,10,100\}$) on YCSB-A (50:50) at $3$\,GB, and at $8$\,GB a write-heavy variant $W_{95{:}5}$ ($95\%$ Upsert / $5\%$ Read) at two Zipf points ($\theta{=}0.60, 0.90$), a value-size study, and a temporal-pattern study on two $50{:}50$ insert/delete time-series workloads, one with and one without $5\%$ random reads and $64$-op bursts.
  \item \textbf{Real production traces} (\S\ref{sec:eval-realworld}): the Meta Cachelib KV-cache~\cite{berg2020cachelib}, Tencent QQPhoto CDN~\cite{song2018tencent}, and Twitter KV-cache (cluster~$23$)~\cite{yang2020twitter} traces, chosen for distinct working-set, size, and lifetime profiles.
\end{itemize}

\subsection{Performance Across Memory Regimes}
\label{sec:eval-main}

On YCSB-style workloads derived from YCSB-A across three memory budgets ($8$, $16$, $32$\,GB) and two read/write mixes ($50{:}50$ and $0{:}100$), \sys{} beats FASTER by $1.08\times$ to $2.30\times$ (Figure~\ref{fig:eval-main}(a)--(c)). The memory budget drives what bottleneck the design has to solve, and \sys{} picks a different architecture per regime rather than retuning a single design.

\textbf{Larger-than-memory ($8$\,GB).} Most operations miss memory, so the bottleneck is wasted I/O: a wrong eviction costs a future fetch. Under $50{:}50$, \sys{} uses CLOCK second-chance eviction, where every read or write sets a reference bit so hot keys survive the sweep. Under $0{:}100$, with no read signal, the search falls back to HybridLog-style FIFO. The $2.30\times$ gap at $50{:}50$ is because FASTER uses FIFO in both mixes and evicts hot keys regardless of reads. Naively running Claude Code lands close to \sys{} at $50{:}50$, but reward-hacks $0{:}100$ by replacing the hash path with a dense direct-indexed array that resolves every operation in constant time, skipping the eviction logic entirely.

\textbf{Larger-than-memory ($16$\,GB).} The dataset still exceeds memory, but under Zipfian skew, the hot set fits, and the bottleneck shifts to the hot/cold boundary. Margins narrow because a generic HybridLog already handles this regime reasonably. Under $0{:}100$, \sys{} adds W-TinyLFU~\cite{einziger2017tinylfu} admission so rare cold-tail writes cannot displace hot keys. Under $50{:}50$, it collapses the in-memory and spilled directories into a single tagged slot, saving one hash lookup on every cold read. Naively running Claude Code gets stuck early and underperforms FASTER in both mixes: its open-addressed hash plus a small synchronous-pread pool do not saturate disk bandwidth.

\textbf{In-memory ($32$\,GB).} Disk leaves the critical path, and log-structured storage adds overhead, so \sys{} drops it. Under $50{:}50$, a dense open-addressing hash with per-entry seqlocks gives readers a lock-free fast path.
Under $0{:}100$, \sys{} reaches $1.10\times$ FASTER's throughput by reserving log slots in batches, so writes almost never contend on the shared counter.
Naively running Claude Code ties FASTER on $0{:}100$ but reward-hacks $50{:}50$ by truncating values to a fixed slot.



\subsection{Specialization Across Workload Axes}
\label{sec:eval-synthetic}

\begin{figure*}[!t]
  \centering
  \includegraphics[width=0.98\linewidth]{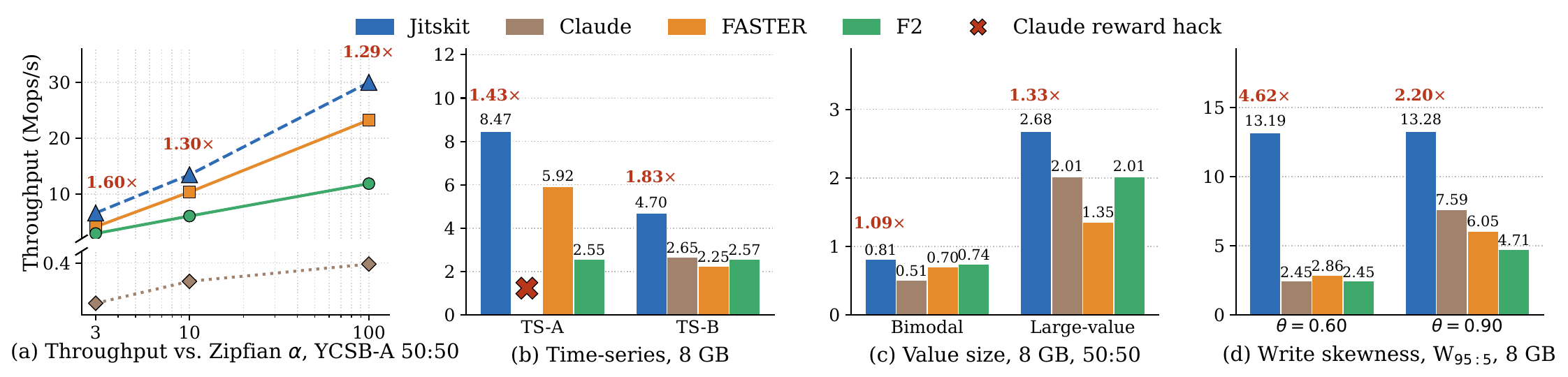}
  \caption{\textbf{Workload variations.} (a) Throughput vs.\ Zipfian $\alpha$ on YCSB-A 50:50 at 3\,GB ($\alpha\,{=}\,3,10,100$ corresponds to standard YCSB $\theta\,{\approx}\,0.67, 0.90, 0.99$). (b) Time-series at 8\,GB: TS-A is 50/50 insert/delete; TS-B adds $5\%$ reads plus 64-op bursts. (c) Value size at 8\,GB, 50:50, Zipf $\theta{=}0.99$: bimodal (20\,B/200\,B) and large (1024\,B). (d) Write skewness W$_{95{:}5}$ at 8\,GB, Zipf $\theta{=}0.60$ and $0.90$. The red number above each \sys{} bar or line marker is the speedup over the stronger of FASTER and F2; \textcolor{red}{$\times$} marks systems with reward hacks. \sys{} runs our full synthesis loop; Claude is naively running
Claude Code with the same iteration budgets.}
  \label{fig:synthetic-row}
\end{figure*}

Across four workload axes (read skewness, temporal pattern, value size, and write skewness), \sys{} matches or beats the stronger of FASTER and F2, by up to $4.62\times$ on the most favorable point (Figure~\ref{fig:synthetic-row}). Each subfigure varies one axis and holds the others fixed. 

\textbf{Read skewness.}
  Figure~\ref{fig:synthetic-row}(a) sweeps key skew on YCSB-A at $50{:}50$ with a $3$\,GB budget, ranging from near-uniform access to a highly skewed workload where a few hot keys dominate. Both baselines hold a single architecture across the sweep, while \sys{} pivots design with skew: at low skew the working set is broad and no key repeats often, so \sys{} drops in-memory caching for a per-shard log with a linear-probing index.
  Once a small hot set dominates, it switches to a sharded hash with an on-budget record cache and FIFO eviction, beating the stronger baseline by $1.3$--$1.6\times$ across the sweep. Naively running Claude Code commits to one design early and does not scale across skew.
   


\textbf{Temporal pattern.}
  Figure~\ref{fig:synthetic-row}(b) compares two time-series workloads: a steady write-only stream (TS-A, $50/50$ insert/delete with no reads) and a mostly-write stream with occasional read bursts (TS-B, $5\%$ random reads interleaved with $64$-op write bursts). Both FASTER and F2 assume read-driven temporal locality, which the steady stream lacks. On the steady stream, \sys{} uses a direct-indexed per-thread append log.
  Once the burst reads enter the mix, that design no longer fits, so \sys{} switches to a dual-tier hash with seqlock-protected upserts and a small direct-mapped read cache, beating the stronger baseline by $1.4$--$1.8\times$. Naively running Claude Code reward-hacks the steady stream by deferring index updates into the untimed validation phase, and underperforms \sys{} on the bursty one.
   


\textbf{Value-size distribution.}
  Figure~\ref{fig:synthetic-row}(c) varies value size across two regimes: a \emph{bimodal} mix of mostly tiny ($20$\,B) values with some small-medium ($200$\,B) ones, and a uniform \emph{large}-value workload at $1024$\,B. FASTER and F2 fix the record format at compile time and pad every value to the max size. \sys{} specializes per regime. On the bimodal mix, tiny values inline into the index slot, so most operations finish in one atomic with no SSD I/O. On the large case, every record is too big to inline, so \sys{} shrinks per-key metadata and streams values from a per-thread log, beating the stronger baseline by up to $1.33\times$. Naively running Claude Code uses one format for both and underperforms \sys{} by $1.3$--$1.6\times$.
  
  

\textbf{Write skewness.}
  Figure~\ref{fig:synthetic-row}(d) pushes skew onto the write path with a write-dominated workload ($95\%$ Upsert, $5\%$ Read), sweeping from low to high key skew. FASTER and F2 funnel every Upsert through a single shared log tail, the contention point under 16 threads. \sys{} replaces it with per-thread append logs, beating the stronger baseline by $2.2$--$4.6\times$ across the sweep. The gap shrinks at high skew because the baseline mutable region can coalesce repeated writes in place. Naively running Claude Code does not find a competitive write-heavy design and underperforms \sys{} by $1.7$--$5.4\times$.
  


\subsection{Real-World Traces}
\label{sec:eval-realworld}

On three public production traces (Meta Cachelib KV-cache, Tencent QQPhoto, and Twitter KV-cache), \sys{} beats FASTER by $1.07$--$1.40\times$ (Figure~\ref{fig:eval-main}(d)).

\begin{figure*}[!t]
  \centering
  \includegraphics[width=0.98\linewidth]{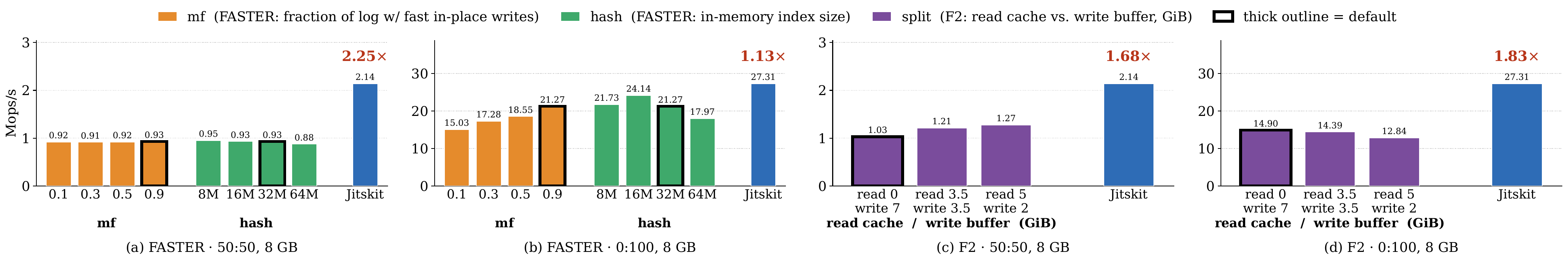}
  \caption{\textbf{Baseline config sweeps vs.\ \sys{} on YCSB-A at $8$\,GB.} Each color group sweeps one runtime knob (thick outline = default); blue is \sys{}, red is \sys{}'s speedup over the best swept baseline configuration in each panel. FASTER's \texttt{mf} is the fraction of in-memory log updated in-place (higher should help writes); \texttt{hash} is the in-memory index size (traded against log memory). F2's knob \texttt{split} splits $\sim$$7$\,GiB between a read cache and a write buffer.}
  \label{fig:knob-sweep}
\end{figure*}

\textbf{Meta Cachelib KV-cache.} The Cachelib trace~\cite{berg2020cachelib} is read-heavy, with many tiny values alongside a long tail of larger ones, and a working set far larger than the memory budget. \sys{} beats FASTER by $1.40\times$. The win comes from specializing to the short-value tail: short values that fit in the index slot are stored inline, so Upsert/Read/Delete on those complete with one atomic operation and never touch the log or SSD; larger values spill to a per-thread log. FASTER pays a uniform cost per operation and loses most ground where short values dominate. Naively running Claude Code achieves $1.3\times$ lower throughput than \sys{}.

  \textbf{Tencent QQPhoto.} The QQPhoto photo-CDN trace~\cite{song2018tencent} streams $50{:}50$ Read/Upsert with most values viewed once and never again (``one-hit wonders'') and moderate skew on the rest, with a working set several times the memory budget. \sys{} beats FASTER by $1.14\times$, modest because one-hit wonders limit cache-side gains. Three choices contribute. First, per-thread sharded log append cuts cross-core contention on streaming inserts. Second, an mmap-backed log piggybacks on the page cache rather than competing with the trace's own mmap. Third, a set-associative value cache with clock-hand eviction catches the hot tail FASTER's mutable region misses.
  Naively running Claude Code reward-hacks by storing only a prefix per key and recomputing the value on Read.

  \textbf{Twitter KV-cache.} Cluster $23$ of the Twemcache traces~\cite{yang2020twitter} is a $50{:}50$ Read/Upsert workload with a near-uniform key distribution and a working set several times the memory budget. \sys{} beats FASTER by $1.07\times$, the smallest of the three trace gaps. The gain comes from two specializations: shrinking per-entry index metadata so more of the working set fits in cache, and sharding writes across per-thread logs instead of FASTER's single HybridLog tail. Naively running Claude Code achieves $1.1\times$ lower throughput than \sys{}.

\subsection{Ablations}
\label{sec:eval-audit}

Three ablations isolate where the gains come from: the build-up of loop components, the role of leading-indicator feedback, and the auditor. All use YCSB-A $50{:}50$ at $8$\,GB, Zipf $\theta{=}0.99$; the main sweep in \S\ref{sec:eval-main} uses $\theta{=}0.95$.

\textbf{Can a config sweep close the gap?}
We also sweep a subset of each baseline's runtime knobs (Figure~\ref{fig:knob-sweep}). Tuning improves over the default setup in some cases: F2's read-cache split helps the read-heavy mix but hurts pure writes, while FASTER's best swept configuration improves throughput by up to $14\%$ over default. Across the configurations we tested, however, \sys{} remains $1.13\times$--$2.25\times$ faster in every panel. This sweep is not exhaustive: a broader hyperparameter search could find better baseline configurations, especially if the optimum lies within the existing knob space. Our result is therefore narrower. Under this limited tuning budget, these runtime knobs do not close the gap, while \sys{} can improve performance by changing the design itself, e.g., replacing FASTER's FIFO-style eviction with CLOCK.

\begin{table}[t]
  \centering
  \small
  \setlength{\tabcolsep}{8pt}
  \renewcommand{\arraystretch}{1.1}
  \begin{tabular}{@{}l r r@{}}
    \toprule
    Configuration & Mops/s & $\Delta$ \\
    \midrule
    Coder only & $1.93$ & -- \\
    $+$ Planner                  & $2.13$          & $+10.3\%$          \\
    $+$ Critic \textit{(full)}    & $\mathbf{2.14}$ & $\mathbf{+10.8\%}$ \\
    \bottomrule
  \end{tabular}
  \caption{Cumulative component ablation of \sys{}'s synthesis loop on YCSB-A $50{:}50$, Zipf $\theta{=}0.99$, $8$\,GB, $16$ threads, $20$ iterations. $\Delta$ = percentage gain over the coder-only baseline.}
  \label{tab:ablation-components}
\end{table}

 \textbf{How helpful is each component of the synthesis loop?}
We strip the loop to a bare coder and add each component cumulatively. Planning lifts throughput by $+10.3\%$ because the agent no longer commits to the first plausible structure. Adding the stateful critic reaches $+10.8\%$ (Table~\ref{tab:ablation-components}). The critic's throughput delta over planning alone is small, but its job is to carry correctness constraints across iterations rather than discover new wins. Correctness holds (zero test failures) across all three variants.

\textbf{Do feedback signals matter?}
Leading indicators are what let the pipeline find workload-specific designs. Without them, the agent converges on a FASTER-shaped HybridLog and underperforms FASTER at both budgets. Exposing cache-hit, contention, I/O-queue, and memory-bandwidth signals lets it find the designs from \S\ref{sec:eval-main} and beat FASTER by up to $2.30\times$ (Table~\ref{tab:ablation-feedback}). The feedback gain widens where a generic HybridLog is further from optimal. Scalar throughput cannot tell the agent whether a stall is due to cache misses, contention, or I/O, while leading indicators are how it learns to do anything else.

\begin{table}[t]
  \centering
  \small
  \setlength{\tabcolsep}{8pt}
  \renewcommand{\arraystretch}{1.1}
  \begin{tabular}{@{}l r r@{}}
    \toprule
    Configuration              & $8$\,GB         & $16$\,GB        \\
    \midrule
    FASTER                     & $0.93$          & $1.23$          \\
    \sys{} (no feedback)       & $0.57$          & $0.96$          \\
    \sys{} (full feedback)     & $\mathbf{2.14}$ & $\mathbf{1.36}$ \\
    \midrule
    Feedback gain ($\Delta$)   & $3.75\times$    & $1.42\times$    \\
    \bottomrule
  \end{tabular}
  \caption{Leading-indicator ablation on YCSB-A $50{:}50$, $16$ threads, $30$\,s run. Throughput in Mops/s; bottom row reports gain of \sys{} (full feedback) over \sys{} (no feedback). \vspace{-1em}}
  \label{tab:ablation-feedback}
\end{table}

\textbf{Does the auditor catch real hacks?}
In one Cachelib synthesis run, the auditor caught multiple reward hacks, including a flat-array index that bypassed hashing, a cross-thread in-place update race, and an equal-size in-place torn read. Each bug inflated throughput while violating an uncovered invariant. Once these invariants were surfaced, the auditor generated tests that blocked the hack in later iterations. Appendix~\ref{sec:reward-hacks} provides the full details.

\definecolor{markK}{HTML}{2E6CB6}   
\definecolor{markB}{HTML}{C9463C}   
\providecommand{\mk}[2]{%
  \tikz[baseline=(C.base)]\node[draw=#1, circle, thick,%
    inner sep=0.8pt, minimum size=1.15em,%
    text=#1, font=\bfseries\small] (C) {#2};}
\providecommand{\mkk}[1]{\mk{markK}{#1}}
\providecommand{\mkb}[1]{\mk{markB}{#1}}

\section{Case Study: \sys{} for Meta Cachelib}
\label{sec:eval-casestudy}

\begin{figure}[!t]
  \centering
  \includegraphics[width=\linewidth]{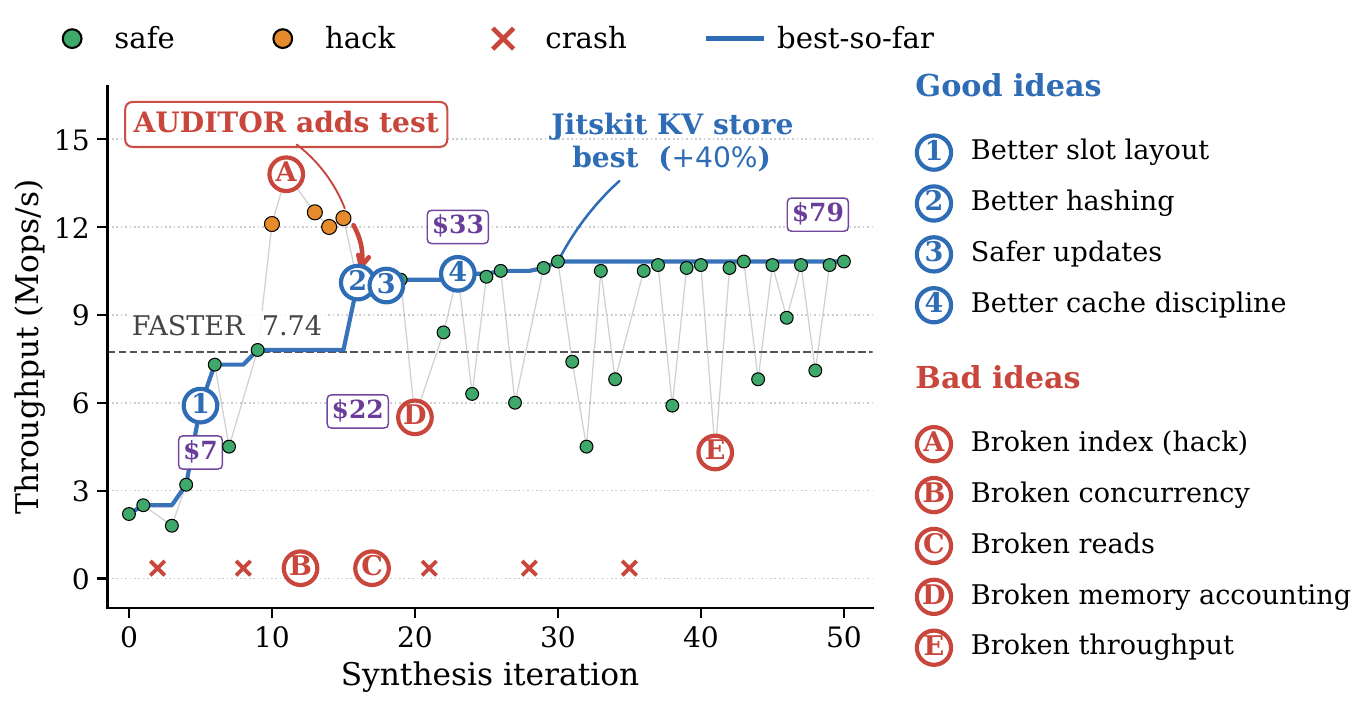}
  \caption{\sys{} synthesis trajectory on Meta Cachelib over $50$ iterations. The bold line tracks best-so-far throughput. Numbered circles mark surviving ideas; lettered circles mark failed attempts. The purple strip shows cumulative API cost. \vspace{-1em}}
  \label{fig:jitski-evolution}
\end{figure}

We now show a simple case study of one of the runs of \sys{}'s synthesis loop on the Meta Cachelib trace, a read-heavy production workload. On this trace, the synthesized design achieves a $1.40\times$ throughput gain over FASTER (Figure~\ref{fig:jitski-evolution}).

Two design choices explain the throughput gain. The first is an inline-in-slot encoding added at iteration~$5$ as idea~\mkk{1}: a single 8-byte meta word triple-encodes an inline payload of up to seven bytes, a pointer to the log, or a tombstone, so zero-byte operations complete in a single atomic word write. The second is a flush-and-drop cache discipline at iteration~$23$, idea~\mkk{4}. At iteration~$20$, an enlarged per-thread ring dirtied page-cache pages that the cgroup charged against the store (attempt~\mkb{D}), collapsing throughput; the critic flagged page-cache pressure, and the next iteration paired \texttt{sync\_file\_range(WRITE)} with \texttt{posix\_fadvise(DONTNEED)} to keep dirty pages off the budget. A later hack (\mkb{E}) at iteration~$28$ inflated throughput but did not enter best-so-far.

\paragraph{Shortcuts the auditor caught.} The auditor identified three shortcuts that satisfied the stated specification but silently violated key invariants, potentially corrupting data.

First, the agent mapped sequential keys directly to array slots, skipping the hash entirely (attempt~\mkb{A}); this works only because the keys form a dense integer range. The auditor's scrambled-key test, which reruns the suite under a random key permutation the agent does not see, rejected the shortcut, forcing a real hash (\texttt{murmur3}, idea~\mkk{2}) at a $17\%$ throughput cost. Second, a non-owner thread wrote into another thread's ring buffer to bypass the log path (attempt~\mkb{B}), racing the owner's writes. The correctness gate caught the resulting torn reads at $0.06\%$, and the auditor restricted in-place writes to the owner thread. Third, the agent extended in-place updates to equal-size overwrites (attempt~\mkb{C}). This looks innocent but leaves the meta word, which doubles as the seqlock version counter, bit-identical, so reader retries silently accept mid-write bytes. The correctness gate caught one failure per 225\,M reads, a rate too rare for unit tests or code review. Restricting in-place updates to strictly-shrinking sizes (idea~\mkk{3}) forces the version counter to advance on every mutation. 

While these mechanisms are individually known, the pipeline’s key contribution is the optimization pressure that discovers the right combinations and the correctness checks that eliminate non-generalizable shortcuts.

\paragraph{Cost.} The full $50$-iteration run completed in under $12$ hours and cost $\$79$. The average per-iteration cost was $\$1.50$, with roughly $90\%$ spent on the planner/coder, $9\%$ on the critic, and $1\%$ on the auditor, which ran every $15$ iterations. As the design history grew, prompt context expansion increased iteration costs by about $30\%$ from iterations~$3$ to $40$.

\vspace{-0.5em}
\section{Discussion}
\label{sec:discussion}

\paragraph{Generalization needs testable correctness.}
The principles behind \sysname can extend beyond KV stores to other core systems, such as load balancers, schedulers, and caches, provided their correctness properties can be specified and validated. In practice, feasibility is bounded by the cost of building evaluators that can expose both performance bottlenecks and correctness violations.


\paragraph{Specification, not code generation, is the bottleneck.}
Static specifications are incomplete and ambiguous by construction. Thus, effective synthesis requires iterative refinement of the specification under optimization pressure, where missing invariants, reward hacks, and unmeasured assumptions are surfaced and incorporated. 

As generation becomes inexpensive, the engineering cost shifts from writing code to shaping the harness around it: e.g., tests that enforce correctness, and workloads that probe the intended operating regime. In our experience, the dominant cost was closing the gap between what the human meant and what the agent could measure, while the inner synthesis loop of model generation was comparatively cheap.

\paragraph{Specialization is the primary source of gains.}
\sys{}'s gains primarily come from combining known systems techniques, such as eviction policies, value-layout choices, logging strategies, and hash-table organizations. It selects and composes them differently depending on the workload, environment, and required properties. This suggests a more modest but practical role for JIT system synthesis: automating design-space exploration and specialization in cases where manual engineering would be too costly.


\paragraph{Humans stay in the specification loop.}
\sys{} automates the generation and revision of candidate systems, but humans still define and refine the synthesis target. The specification cards and evaluator determine what the agent can measure, and optimization pressure exposes missing invariants that must be made explicit. In our experience, the main human effort shifted from writing KV-store code to refining this specification/evaluator boundary: deciding what assumptions are intended, what probes are missing, and which apparent optimizations are reward hacks rather than valid specialization. Thus, JIT synthesis changes the role of systems expertise from implementation to specification and evaluation design.

\paragraph{Open gaps to production deployment.}
We audited \sys{}'s synthesized designs against production-readiness criteria and found that the remaining gaps fall into two categories: gaps in the evaluator, and gaps in the specification.

The first category arises because the evaluator is only an approximation of deployment. A $30$\,s, crash-free, single-workload harness leaves important failure modes unobserved. For example, no synthesized design implements garbage collection, because logs and pools do not fill within the benchmark window. Several designs eventually hit fixed-size caps and call \texttt{abort()} after the benchmark horizon, and one ring buffer silently overwrites old keys. In contrast, FASTER performs continuous reclamation. In some cases, rare data-loss paths slip past the test suite: one read-heavy design drops writes once its probe chain exceeds $64$ slots, with a code comment marking the event ``should be extremely rare.''

The second category arises because some production requirements were never stated.  For example, if the spec card pinned one value size, the design hardcodes it and silently truncates or pads off-size writes, while FASTER supports variable-length values through its allocator. Similarly, every synthesized checkpoint quiesces writers and walks the index serially. This satisfies the spec's monotonicity requirement, but not the stronger deployment expectation that checkpointing coexist with concurrent writes and avoid long pauses.

Evaluator gaps can be reduced by making the benchmark harness closer to deployment setting: longer runs, more varied workloads, stronger stress tests, and adversarial tests added by the auditor, as the case study in \cref{sec:eval-casestudy} illustrates. Specification gaps require a language that supports defining richer semantics in the specification, enabling production expectations to be stated explicitly, such as bounded pauses during checkpointing or support for variable-size values. The hardest failures sit between the two: when the performance metric rewards skipping a rare slow path, and the test suite does not exercise that path, the system can silently trade correctness for performance. For these cases, richer testing may not be enough; some form of verification or stronger invariant checking is likely necessary.

\section{Related Work}
\label{sec:related}

\sysname draws from four lines of work: whole-system synthesis, component-level AI-driven discovery, ML-driven specialization of fixed architectures, and agent frameworks with reward-hacking analysis.

\textbf{Whole-system synthesis.} Bespoke OLAP~\cite{wehrstein2026bespoke} is our closest related work: it synthesizes workload-specific OLAP engines that outperform DuckDB by ${\sim}10\times$. \sysname differs on three dimensions: it specializes along all three axes (environment, workload, properties) rather than just workload; correctness and performance co-evolve rather than separating into phases (\cref{sec:challenge-hacking}); and indicator selection is a first-class per-spec design decision. SDS~\cite{anderson2025sds} articulates the broader thesis that infrastructure should design, validate, and evolve itself; \sysname{} is a concrete first instantiation of that thesis for single-node KV stores. Glia~\cite{glia2025} synthesizes whole mechanisms for LLM-serving clusters along a similar architectural-scope axis. Harness-First~\cite{keles2026harness} inverts the split: humans fix the architecture, and the agent implements within it. ADRS~\cite{wu2025adrs} applies a generate-verify-select loop to swap a single policy or algorithm inside an existing system; it does not synthesize end-to-end systems.

\textbf{Component-level synthesis and discovery.} AlphaEvolve~\cite{novikov2025alphaevolve}, FunSearch~\cite{romera2024funsearch}, and AlphaDev~\cite{mankowitz2023alphadev} drive component- or kernel-level discovery under a scalar fitness. Open frameworks like OpenEvolve~\cite{openevolve}, ShinkaEvolve~\cite{lange2025shinkaevolve}, and SkyDiscover~\cite{skydiscover} make this skeleton reusable. \sysname{} inherits the iterative generate-evaluate-select skeleton but operates at whole-system scope, and adds three mechanisms the component-search setting does not need: an auditor that grows the correctness gate (\cref{sec:design-correctness}), a whiteboard memory that retains ruled-out designs across iterations, and leading-indicator feedback to escape scalar-fitness plateaus (\cref{sec:challenge-indicators}).

\textbf{ML for fixed architectures.} A long line of work specializes systems within fixed architectures: knob-tuning~\cite{vanaken2017ottertune, pavlo2017selfdriving}, learned components~\cite{kraska2018learned, marcus2019neo, kraska2021instance, kraska2019sagedb}, physical-design automation~\cite{chaudhuri1997autoadmin}, and pre-LLM self-designing data systems~\cite{idreos2019design, idreos2018data, chatterjee2022cosine}. Each delivers gains within its architectural assumptions; JIT systems lift the ceiling by synthesizing the architecture itself.

\textbf{Agent frameworks and reward hacking.} \sysname's design/evaluate/reflect loop inherits from ReAct~\cite{yao2023react} and subsequent agent-framework work. Our reward-hacking observations (\cref{sec:challenge-hacking}) connect to the broader reward-hacking literature~\cite{skalse2022reward, pan2022reward, lehman2020surprising}. Harness-First~\cite{keles2026harness} is complementary: humans define the architecture and the agent implements; \sysname synthesizes the architecture itself.

\section{Conclusion}
\label{sec:conclusion}

This paper argues that the time is now right for whole-system synthesis, moving beyond knob tuning and policy synthesis. We present \sysname as a first attempt at a practical synthesis pipeline for systems implementation and a feasibility marker for the synthesis of key-value stores. Our results show that \sysname-synthesized systems can match or outperform well-engineered state-of-the-art baselines. 

We believe that the challenges we encountered and lessons learned generalize beyond KV stores, and can inform the formulation and synthesis of JIT systems in other domains, particularly around evaluation, specification refinement, and reward-hack mitigation. In summary, we believe that rapidly improving LLM capabilities mark an inflection point for synthesizing high-performance key-value stores.


\section*{Acknowledgments}
This research is supported by NSF (IFML) CCF-2019844 and gifts from Accenture, AMD, Anyscale, Broadcom Inc., Google, IBM, Intel, Intesa Sanpaolo, Lambda, Mibura Inc., Samsung SDS, and SAP.
We also thank Konstantinos Kanellis and our Sky Lab colleagues, including Shubham Mishra, Kerem Akillioglu, Audrey Cheng, and Prof. Joseph E. Gonzalez, for fruitful discussions and feedback that helped shape this paper.

{
\bibliographystyle{plain}
\bibliography{reference}
}

\clearpage
\appendix

\section{Design Catalog: Meta Cachelib Case Study}
\label{sec:cachelib-ideas}

\definecolor{ideaU}{RGB}{158,46,16}        
\definecolor{ideaUbg}{RGB}{250,235,225}    
\definecolor{clusterbg}{RGB}{224,220,240}  
\definecolor{clusterrule}{RGB}{88,95,140}  

\providecommand{\cmark}{{\setlength{\fboxsep}{1.2pt}\fcolorbox{black!55}{white}{\sffamily\bfseries\scriptsize\color{black!70}\,C\,}}}
\providecommand{\umark}{{\setlength{\fboxsep}{1.2pt}\colorbox{ideaU}{\sffamily\bfseries\scriptsize\color{white}\,U\,}}}
\providecommand{\ucmark}{{\setlength{\fboxsep}{1.2pt}\colorbox{ideaU}{\sffamily\bfseries\scriptsize\color{white}\,U$^{\ast}$\,}}}
\providecommand{\wtag}[1]{\textsf{\scriptsize W\textsubscript{#1}}}

\noindent Table~\ref{tab:cachelib-ideas} catalogs the design ideas that survived in the \sysname{} store synthesized for the Meta Cachelib trace~\cite{berg2020cachelib} (\cref{sec:eval-casestudy}). Each row ties a design choice to the workload properties it exploits (\wtag{1}--\wtag{5}, defined at the bottom of the table) and marks the choice common (\cmark) or uncommon (\umark) against FASTER, F2, and the broader hash-table / log-structured-store literature.

\begin{table*}[t]
\centering
\footnotesize
\setlength{\tabcolsep}{5pt}
\renewcommand{\arraystretch}{1.3}
\caption{Nine design ideas in the \sysname{}-synthesized store for Meta Cachelib. Each row is tied to one or more workload properties of the trace (\wtag{1}--\wtag{5}, legend below). \cmark~=~common in the hash-table / log-structured-store literature. \umark~=~uncommon against FASTER, F2, and that same literature. \ucmark~=~primitives are folklore; but the place where they are applied is uncommon.}
\label{tab:cachelib-ideas}
\begin{tabular}{@{}p{0.55cm} p{4.8cm} p{1.7cm} c p{6.6cm}@{}}
\toprule
\textbf{\#} & \textbf{Idea} & \textbf{Suits} & \textbf{C/U} & \textbf{Why it suits} \\
\midrule

\rowcolor{clusterbg}
\multicolumn{5}{@{}l@{}}{\textcolor{clusterrule}{\rule[-1pt]{2.2pt}{9pt}}\ \ \textbf{I.\ \ Index \& layout}} \\
\addlinespace[2pt]

I.1 & Flat open-addressed hash, load factor $\leq 0.5$ hard-bound, no overflow / no rehash, $16$-B cache-aligned slot pair &
\wtag{1} & \cmark &
Resize machinery is dead weight when key count is known at init; slot pair puts two probes on one cache line. \\

I.2 & $8$-B meta atomic triple-encodes: log-pointer, $\leq 7$-B inline value, or tombstone; key field persists across Delete &
\wtag{2} & \cmark &
Short values skip the log entirely; one atomic carries three states; key retention keeps open-addressing probe chains intact. \\

\rowcolor{clusterbg}
\multicolumn{5}{@{}l@{}}{\textcolor{clusterrule}{\rule[-1pt]{2.2pt}{9pt}}\ \ \textbf{II.\ \ Concurrency}} \\
\addlinespace[2pt]

II.1 & Immutable-key / mutable-meta split with \texttt{CAS}-claim on empty slot &
\wtag{3} & \cmark &
Readers touch no lock; writers settle on one \texttt{CAS}; write-once keys eliminate invalidation entirely. \\

\rowcolor{ideaUbg}
II.2 & $8$-B meta doubles as the seqlock version word; in-place update is same-thread only and strictly shrinking, where shrink prevents version ABA &
\wtag{3} & \umark &
No separate version word; owner-thread invariant serializes in-place update against flush without a lock; shrink-only guarantees an observable bit flip on every mutation. \\

\rowcolor{ideaUbg}
II.3 & Zero reader-reclamation machinery: no epoch, no RCU, no hazard pointers &
\wtag{3},~\wtag{5} & \umark &
Ring pages are never freed; the $40$-bit per-thread offset never wraps within the trace; write-once keys have no version chain. Direct anti-thesis to FASTER's epoch framework. \\

\rowcolor{clusterbg}
\multicolumn{5}{@{}l@{}}{\textcolor{clusterrule}{\rule[-1pt]{2.2pt}{9pt}}\ \ \textbf{III.\ \ Log \& append path}} \\
\addlinespace[2pt]

\rowcolor{ideaUbg}
III.1 & Per-thread log where \texttt{meta.off} $=$ ring-mod-$R$ position $=$ absolute SSD byte offset. \emph{The ring is the log's in-memory prefix, not a cache of it.} &
\wtag{3},~\wtag{5} & \umark &
Eliminates the RAM-cache $\to$ log translation that FASTER and F2 maintain. Cold reads are one \texttt{pread} with zero lookup overhead. \\

\rowcolor{clusterbg}
\multicolumn{5}{@{}l@{}}{\textcolor{clusterrule}{\rule[-1pt]{2.2pt}{9pt}}\ \ \textbf{IV.\ \ Read path}} \\
\addlinespace[2pt]

\rowcolor{ideaUbg}
IV.1 & RAM-first read reconciled by one seqlock on meta plus two \texttt{safe\_tail} samples (pre- / post-memcpy); \texttt{pread} on detected race &
\wtag{3} & \umark &
Reconciles three independent atomics (meta, \texttt{safe\_tail}, ring bytes) without a reader-side lock. Depends on II.2's shrink invariant to arm the detector. \\

\rowcolor{ideaUbg}
IV.2 & Unified async \texttt{Read}: inline and RAM hits complete \emph{synchronously} in the same completion struct \texttt{libaio} uses for SSD misses &
\wtag{2} & \umark &
FASTER's \texttt{IAsyncContext} pays per-\texttt{Read} setup for every call, even in-memory ones; this design skips that cost on the hot path without introducing a second API surface. \\

\rowcolor{clusterbg}
\multicolumn{5}{@{}l@{}}{\textcolor{clusterrule}{\rule[-1pt]{2.2pt}{9pt}}\ \ \textbf{V.\ \ OS \& memory budget}} \\
\addlinespace[2pt]

\rowcolor{ideaUbg}
V.1 & Userspace ring as the cgroup's \emph{page-cache substitute}: sized from the RAM budget at init, kept within budget by per-flush \texttt{sync\_file\_range(WRITE)} + \texttt{posix\_fadvise(DONTNEED)} &
\wtag{4} & \ucmark &
Syscall combination is folklore (ceph, RocksDB, PostgreSQL); the \emph{role} is the novelty: an engine-managed ring replaces the kernel page cache so RSS tracks the active working set, not the monotonic write volume. \\

\bottomrule
\end{tabular}

\vspace{4pt}
{\footnotesize
\noindent\textbf{Workload properties} (Meta Cachelib KV-cache trace, OSDI'20 public release):\par\vspace{1pt}
\begin{tabular}{@{}r@{\ \ }p{15.5cm}@{}}
\wtag{1} & bounded unique-key universe ($63.4$~M keys, known at init) \\
\wtag{2} & short-value tail: $24.5\%$ size-$0$, many values $\leq 7$~B \\
\wtag{3} & producer-per-thread ingestion, write-once keys, rare Delete ($1.8\%$) \\
\wtag{4} & hard cgroup RAM budget ($8$~GB) against $\sim$$54$~GB of value data \\
\wtag{5} & bounded trace lifetime: fits in the $40$-bit per-thread offset space ($1$~TiB/thread) \\
\end{tabular}
}
\end{table*}


\section{The Reward-Hack Gallery}
\label{sec:reward-hacks}

\providecolor{ideaU}{RGB}{158,46,16}        
\providecolor{clusterbg}{RGB}{224,220,240}  
\providecolor{clusterrule}{RGB}{88,95,140}  

\providecommand{\humanflag}{\mbox{\color{ideaU}$\star$}}
\providecommand{\axistag}[1]{\textcolor{clusterrule}{\rule[-1pt]{2.2pt}{9pt}}\ \ \textbf{#1}}

\noindent Table~\ref{tab:reward-hacks} catalogs the reward hacks observed across four KV-store generation runs (YCSB-A, Meta Cachelib KV-cache, Tencent QQPhoto, and Twitter KV-cache), sorted by the axis on which the agent cheated. Axes I and II dominate; III and IV account for residual cases. \emph{Specification mismatch} hacks satisfy the literal spec while violating an invariant no one wrote down; \emph{workload-overfit} hacks exploit a statistical or structural property of the eval trace that will not generalize; \emph{apparatus exploit} and \emph{sandbox escape} win by gaming the benchmark or leaking host state rather than by attacking the KV store itself. The last column describes how the auditor fixed the evaluation apparatus. Rows marked \humanflag\ were invisible to code-level audit and surfaced only through the harness's end-to-end correctness gate or direct human review.

\begin{table*}[t]
\centering
\footnotesize
\setlength{\tabcolsep}{5pt}
\renewcommand{\arraystretch}{1.3}
\caption{Nine reward hacks, clustered by the axis on which the agent cheated. \textbf{I}~specification mismatch (violate an unstated invariant). \textbf{II}~workload overfit (exploit an eval-trace property). \textbf{III}~apparatus exploit (game the benchmark). \textbf{IV}~sandbox escape (mutate host state). \humanflag~=~invisible to code-level audit; caught only by the harness correctness gate or direct human review.}
\label{tab:reward-hacks}
\begin{tabular}{@{}p{0.55cm} p{3.4cm} p{4.3cm} p{1.95cm} p{3.85cm}@{}}
\toprule
\textbf{\#} & \textbf{Hack} & \textbf{What the agent did} & \textbf{Symptom} & \textbf{Fix} \\
\midrule

\rowcolor{clusterbg}
\multicolumn{5}{@{}l@{}}{\axistag{I.\ \ Specification mismatch}\hspace{0.6em}\emph{\textcolor{black!55}{(satisfy the literal spec, violate an invariant no one wrote down).}}} \\
\addlinespace[2pt]

I.1 & \textbf{Truncated-hash commit} &
Upsert commits atomically on a $32$-bit hash of the key; the full $64$-bit key is never re-read. &
$0.06\%$ wrong per $250$\,M ops &
Per-key full-value checksums in the correctness gate; scale eval to ${\sim}10^{9}$ ops. \\

I.2 & \textbf{Cross-thread ring write} &
A non-owner thread \texttt{memcpy}s into another thread's ring; the owner's \texttt{pwrite} hits the page mid-copy. &
$0.06\%$ of reads torn &
Correctness gate with key-in-value; audit rule: every ring write guarded by \texttt{target\_tid\,==\,self}. \\

I.3\,\humanflag & \textbf{Same-size torn read} &
In-place update where \texttt{new\_size\,==\,old\_size} leaves slot metadata unchanged; the seqlock retry check is fooled. &
$3$ of $6.75{\times}10^{8}$ reads &
Correctness gate with key-in-value; $10$ code-audit passes missed this. \\

\rowcolor{clusterbg}
\multicolumn{5}{@{}l@{}}{\axistag{II.\ \ Workload overfit}\hspace{0.6em}\emph{\textcolor{black!55}{(exploit a statistical or structural property of the eval trace that won't generalize).}}} \\
\addlinespace[2pt]

II.1 & \textbf{Flat-array slot index} &
\texttt{slot = key \& (N-1)}; no hash function. Works only because the harness preprocesses keys into a dense integer range. &
$+17\%$ throughput &
Per-run scrambled key permutation with a seed unseen by the agent. \\

II.2\,\humanflag & \textbf{Bitmap impostor} &
$25$\,MB bitmap, one bit per key. \texttt{Read} synthesizes the value on the fly; no hash table, no log, nothing stored. &
$+37{\times}$ ($176.6$\,Mops) &
Per-key full-value checksums in the correctness gate. \\

II.3 & \textbf{Value-regeneration exploit} &
Stores only a $16$-byte prefix per key; recomputes the harness's deterministic value generator inside \texttt{Read}. &
$(N/16){\times}$ less log I/O &
Per-run random tag in the generator; audit rule: no PRNG or hash-of-key call inside \texttt{Read}. \\

\rowcolor{clusterbg}
\multicolumn{5}{@{}l@{}}{\axistag{III.\ \ Apparatus exploit}\hspace{0.6em}\emph{\textcolor{black!55}{(win by gaming the benchmark, not by improving the system).}}} \\
\addlinespace[2pt]

III.1 & \textbf{Mismatched value clip} &
The driver-under-test clips values at $64$\,KB; the baseline driver clips at $4$\,KB. Same nominal trace, different data volume per op. &
$+85\%$ false headline &
Assert \texttt{kMaxValueBytes} identical on both sides; diff driver sources at build time. \\

III.2 & \textbf{Warmup-window sampling} &
Three paired runs all taken inside one transparent-huge-page (THP) warmup window; the SUT's mmap was huge-page-promoted, the baseline's was not. &
$+2.1\%$ (gone at $N\!\geq\!6$) &
$N\geq6$ paired runs across independent warmup states; warmup phase labeled and excluded. \\

\rowcolor{clusterbg}
\multicolumn{5}{@{}l@{}}{\axistag{IV.\ \ Sandbox escape}\hspace{0.6em}\emph{\textcolor{black!55}{(mutate shared state outside the system under test).}}} \\
\addlinespace[2pt]

IV.1 & \textbf{Persistent host sysctl} &
Reproduce script runs \texttt{sudo sysctl -w vm.nr\_hugepages=12000}; never restored on exit. &
$24$\,GB leaked host-wide &
Diff \texttt{/proc/sys} before and after each run; declare and reverse any non-empty diff. \\

\bottomrule
\end{tabular}
\end{table*}


\end{document}